\documentclass[12pt,reqno]{amsproc}

\usepackage[margin=1in]{geometry}

\usepackage[T1]{fontenc}
\usepackage{bbm}
\usepackage[usenames]{color}
\usepackage{mathpazo}
\usepackage{amsmath,amssymb,amsthm}
\usepackage{wrapfig}
\usepackage{tikz-cd}
\usepackage[shortlabels]{enumitem}

\usepackage{caption}
\usepackage{mathrsfs,bm,nicefrac}
\usepackage[all,cmtip]{xy}
\usepackage{subcaption}
\usepackage{wrapfig}
\usepackage{physics}

\newcommand{\orcid}[1]{\href{https://orcid.org/#1}{\includegraphics[width=10pt]{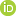}}}

\definecolor{PineGreen}{rgb}{0.0,0.47,0.44}
\definecolor{MidnightBlue}{rgb}{0.1,0.1,0.44}
\definecolor{magenta}{rgb}{1.0,0.0,1.0}
\definecolor{bl1}{HTML}{4479A1}
\definecolor{pur1}{HTML}{52196D}
\definecolor{mag1}{HTML}{2AD0F1}
\definecolor{org1}{rgb}{.92,.39.21}
\definecolor{pur2}{rgb}{.53,.47,.7}
\definecolor{desycyan}{rgb}{0.00,0.68,0.93}
\definecolor{desyorange}{rgb}{0.96,0.52,0.07}
\definecolor{desygray}{rgb}{0.47,0.47,0.47}

\usepackage{hyperref}
\hypersetup{
	colorlinks=true,
	linkcolor=desycyan,
	citecolor=PineGreen,
	filecolor=magenta,
	urlcolor=desyorange
}

\usepackage{cleveref}

\makeatletter
\newcommand{\eqnum}{\refstepcounter{equation}\textup{\tagform@{\theequation}}}
\makeatother

\numberwithin{theorem}{section}

\newtheorem*{theorem*}{Theorem}

\theoremstyle{definition}

\theoremstyle{remark}

\AtEndEnvironment{example}{\null\hfill\ensuremath{\diamond}}

\newcommand{\cA}{\mathcal{A}}

\newcommand{\cL}{\mathcal{L}}

\newcommand{\slashed}{\not\!}
\newcommand{\gmunu}{g_{\mu\nu}}
\newcommand{\munu}{{\mu\nu}}

\makeatletter
\DeclareRobustCommand
\myvdots{\vbox{\baselineskip4\p@ \lineskiplimit\z@
		\hbox{.}\hbox{.}\hbox{.}}}
\makeatother

\usepackage{tikz}    
\usetikzlibrary{graphs,graphs.standard,calc,automata}

\usepackage[foot]{amsaddr}

\usepackage{tikz-feynman}
\tikzfeynmanset{compat=1.1.0}

\def\DD{D\kern-.7em\raise0.3ex\hbox{\char '55}\kern.33em}

\usepackage{booktabs}
\usepackage{colortbl}
\definecolor{Ftitle}{RGB}{11,46,108}
\colorlet{tableheadcolor}{Ftitle!25} 
\colorlet{tablerowcolor}{gray!10} 
\title{Renormalization of $\boldsymbol{U(1)}$ Gauge Boson Kinetic Mixing}

\author{Felix Forner\orcid{0009-0002-9436-5627}$^1$}
\email{\hyperlink{felix.forner@tum.de}{felix.forner@tum.de}}
\address[1]{Technical University of Munich, TUM School of Natural Sciences,\newline
\hspace*{1.15cm}Physics Department, James-Franck-Stra{\ss}e 1, D-85748 Garching, Germany}

\author{Felix Tellander\orcid{0000-0001-6418-8047}$^{2,3}$}
\email{\hyperlink{felix@tellander.se}{felix@tellander.se}}
\address[2]{Mathematical Institute, University of Oxford, Oxford OX2 6GG, United Kingdom}
\address[3]{School of Mathematics and Hamilton Mathematics Institute, Trinity College,\newline
\hspace*{1.15cm}Dublin 2, Ireland}

\date{\today}

\begin{document}
\vspace*{-2\baselineskip}%
\hspace*{\fill} \mbox{\footnotesize{\textsc{TUM-HEP-1580/25}}}

\maketitle
\begin{abstract}
    Quantum field theories containing fields with the same quantum numbers allow for mixed kinetic terms in the Lagrangian, leading to off-diagonal elements in the tree-level two-point function. After removing the mixing by a field rotation, the off-diagonal UV divergences cannot be subtracted by a counterterm, still one can show that the theory is renormalizable. We study kinetic mixing of $U(1)$ gauge bosons in an extension of QED with a massive "dark" photon at one-loop order. In general covariant $R_\xi$-gauge, the gauge-fixing function naively obstructs the removal of tree-level mixing but we show that these off-diagonal gauge-dependent contributions cancel. We compare two renormalization schemes: one with and one without kinetic mixing, and relate them via a scale-dependent field transformation, showing that the schemes are equivalent.
\end{abstract}
\section{Introduction}

Renormalizability is key for a quantum field theory to be consistent. In many standard cases, the UV divergences can be subtracted through counterterms at the level of each individual amputated Green's function. However, this is not always possible if the fields mix at loop level. Known examples with this feature are theories with several fields $\Phi_i$ with the same quantum numbers. In these cases, one may allow these fields to mix through non-diagonal kinetic terms in the Lagrangian,
\begin{equation*}
    \mathcal{L}\supset \sum\limits_{i< j} \frac{\kappa_{ij}}{2}\partial_\mu\Phi_i\partial^\mu\Phi_j.
\end{equation*}
These terms may be rotated away at tree-level but mixing will reappear at loop-level. Divergences from the off-diagonal two-point Green's functions $G_{ij}$ must be absorbed by a sort of non-diagonal renormalization. Thus, even if one rotates away $\kappa_{ij}$ initially, one might think that it must be reintroduced at tree-level to provide a counterterm that can cancel this divergence. For scalar fields, it was decisively shown to one-loop order in \cite{Bijnens:2018rqw} that $\kappa_{ij}$ can be rotated away at tree-level and that the theory can be consistently renormalized without reintroducing $\kappa_{ij}$. The resolution lies in the basic fact that it is only physical quantities that are in need of renormalization, and the two-point functions themselves are not physical; only their poles and S-matrix elements. 

As is well-known, in theories without kinetic mixing, the reduction formula of Lehmann, Symanzik, and Zimmerman (LSZ) \cite{Lehmann:1954rq} relates $S$-matrix elements directly to amputated, renormalized correlation functions.
It is important to note that these two objects differ in a theory with kinetic mixing, since one has to modify the LSZ theorem to account for mixing on the external legs \cite{Amoros:2001cp} when computing $S$-matrix elements. Then, the off-diagonal pieces of $G_{ij}$ contribute to the $S$-matrix elements, and thus, they do enter the computation of physical quantities. However, it is possible to define a consistent scheme, without reintroducing $\kappa_{ij}$, where these divergences are absorbed in counterterms of the couplings.

In this work, we address this matter in the context of kinetic mixing of two $U(1)$ gauge bosons. We work in general covariant $R_\xi$-gauge and demonstrate to one-loop order that the theory can be renormalized consistently, both in a basis with and without kinetic mixing, and that the results in the two bases match.

On the level of the Lagrangian, removing the off-diagonal kinetic terms produces non-diagonal gauge fixing functions, which still lead to a non-diagonal tree-level propagator.
We show, however, that these gauge-dependent off-diagonal contributions cancel and that the field redefinition is consistent. We work with standard QED with an additional massive, \emph{dark photon}, which does not couple directly to the electron but interacts only through a mixed kinetic term with the photon. This model setup is inspired by popular Standard Model extensions, e.g.  \cite{Okun:1982xi,Holdom:1985ag,Redondo:2008ec,Arias:2012az,Jaeckel:2012mjv,Graham:2015rva,Bauer:2018onh,Fabbrichesi:2020wbt}, where many of the phenomenological consequences of a kinetically mixed massive dark photon are explored, in particular the possibility of dark photon dark matter. To give a non-vanishing gauge-invariant mass to the dark photon, we employ the Stückelberg mechanism \cite{Stueckelberg:1938hvi,Stueckelberg:1938zz}. 

This paper is organized as follows. First, in \cref{sec: Non-minimal model} we introduce the QED extension with kinetic mixing, renormalize it at one-loop order and derive the renormalization group equations (RGEs). In \cref{sec:basis without kin mix} we do the same in a basis without kinetic mixing, where we only renormalize the physical parameters. We relate the two schemes in \cref{sec: comparison} and conclude in \cref{sec: conclusions}.
\section{QED Extension with U(1) Gauge Boson Kinetic Mixing}\label{sec: Non-minimal model}
We consider QED with kinetic mixing between the photon and an additional, massive $U(1)$ gauge boson, which we will refer to as a dark photon. The theory is defined by the (bare) Lagrangian
\begin{equation}
\begin{split}
    \cL=&-\frac{1}{4}(F_{1\,b}^{\mu\nu})^2+i\overline{\psi}_b\gamma_\mu(\partial^\mu+ie_bA_{1\,b}^\mu)\psi_b-m_b\overline{\psi}_b\psi_b-\frac{1}{2\xi_{1\,b}}(\partial_\mu A_{1\,b}^\mu)^2-\frac{\kappa_b}{2}F_{1\,b}^{\mu\nu}F_{2\,b,\mu\nu}\\
    &-\frac{1}{4}(F_{2\,b}^{\mu\nu})^2+\frac{M_b^2}{2}\left(A_{2\,b}^\mu-\frac{1}{M_b}\partial^\mu\phi_b\right)^2-\frac{1}{2\xi_{2\,b}}(\partial_\mu A_{2\,b}^\mu+\xi_{2\,b} M_b\phi_b)^2, \label{eq:original Lagrangian}
\end{split}
\end{equation}
where $A_{1\, b}$ describes the photon, $A_{2\, b}$ the dark photon with mass $M_b$; $\psi_b$ the electron with mass $m_b$, and we define $
    F_{i\, b}^{\mu\nu} = \partial^\mu A_{i\, b}^\nu-\partial^\nu A_{i\, b}^\mu,\ i=1,2,$
as usual. We denote the symmetry group by $U(1)_{\text{QED}} \times U(1)_{\text{dark}}$.
The second line in the above Lagrangian is the St\"uckelberg Lagrangian. The Stückelberg mechanism \cite{Stueckelberg:1938hvi,Stueckelberg:1938zz} allows us to give a mass to the dark photon in a gauge invariant way, by introducing the Stückelberg field $\phi_b$. For the Lagrangian \eqref{eq:original Lagrangian} to be invariant under the gauge transformations
\begin{align}
    A_{1\, b}^\mu(x) \xrightarrow{U(1)_{\text{QED}}} A_{1\, b}^\mu(x) - \frac{1}{e_b} \partial^\mu \alpha_1(x) \quad \mathrm{and}\quad
    A_{2\, b}^\mu(x) \xrightarrow{U(1)_{\text{dark}}} A_{2\, b}^\mu(x) -  \partial^\mu \alpha_2(x),
\end{align}
we require the Stückelberg field to transform trivially under $U(1)_{\text{QED}}$ and as
\begin{align}
    \phi_b(x) &\xrightarrow{U(1)_{\text{dark}}} \phi_b(x) - M_b \alpha_2(x)
\end{align}
under $U(1)_{\text{dark}}$, such that the mass term of the dark photon is invariant. For the gauge fixing term of $A_{2\, b}$ to be invariant as well, we impose for $\alpha_2(x)$ the Klein-Gordon equation
\begin{align}
    (\partial^2 - \xi_{2\, b} M_b^2) \alpha_2(x) = 0.
\end{align}
To see more clearly that the dark photon field $A_{2\, b}$ acquires a mass in this model, note that, up to a total derivative, the Stückelberg Lagrangian may be rewritten as
\begin{equation}
    \cL_{\text{St{\"u}ckelberg}}=-\frac{1}{4}(F_{2\,b}^{\mu\nu})^2-\frac{1}{2\xi_{2\,b}}(\partial_\mu A_{2\,b}^\mu)^2+\frac{M_b^2}{2}(A_{2\,b}^\mu)^2+\frac{1}{2}(\partial_\mu\phi_b)^2-\xi_{2\,b}\frac{M_b^2}{2}\phi_b^2.
\end{equation}
For a more detailed review of the Stückelberg mechanism, see, for example, \cite{Ruegg:2003ps}.

In our Lagrangian in \cref{eq:original Lagrangian}, there is no direct fermion-dark-photon coupling. The only coupling of the dark photon is through the kinetic mixing to the photon, parametrised by $\kappa_b$. We will now renormalize this theory at one-loop and derive the RGEs. First, we compute the gauge boson propagator and full two-point function, which will be matrix-valued due to the mixing.
\subsection{Two-Point Function and Masses}
The tree-level propagator $P$ is found from a fundamental solution to the equations of motion. The equations of motion from the Lagrangian \eqref{eq:original Lagrangian} are
\begin{equation}
    \begin{pmatrix}
        \partial^2g_{\mu\nu}-\left(1-\frac{1}{\xi_{1\, b}}\right)\partial_\mu\partial_\nu & \kappa_b\left(\partial^2g_{\mu\nu}-\partial_\mu\partial_\nu\right)\\ 
        \kappa_b\left(\partial^2g_{\mu\nu}-\partial_\mu\partial_\nu\right) &
        (\partial^2+M_b^2)g_{\mu\nu}-\left(1-\frac{1}{\xi_{2\, b}}\right)\partial_\mu\partial_\nu 
    \end{pmatrix}\begin{pmatrix}
        A_{1\, b}^\mu\\A_{2\, b}^\mu
    \end{pmatrix}=\begin{pmatrix}
        J^{QED}_{b,\nu}\\ 0
    \end{pmatrix},
\end{equation}
where $J^{QED}_{b,\nu}=\overline{\psi}_b(x)\gamma_\nu\psi_b(x)$ is the standard QED current. By Fourier transforming this matrix and inverting it as a block-matrix, we find the matrix-valued propagator\small
\begin{align} \label{eq:matrix prop mixed basis}
    i P=\begin{pmatrix}
        \frac{-i}{p^2} \left(A(p^2) \left( \gmunu - \frac{p_\mu p_\nu }{p^2} \right) + \xi_{1\, b} \frac{p_\mu p_\nu }{p^2}\right)&  \frac{i\kappa_b}{(1-\kappa_b^2)p^2-M_b^2}\left(g^{\mu\nu}-\frac{p^\mu p^\nu}{p^2}\right) \\
 \frac{i\kappa_b}{(1-\kappa_b^2)p^2-M_b^2}\left(g^{\mu\nu}-\frac{p^\mu p^\nu}{p^2}\right) & \frac{-i}{\left(1-\kappa_b^2\right) p^2-M_b^2}\left(g^{\mu\nu}-\frac{ p^{\mu } p^{\nu }\left(1-\xi_{2\, b} \left(1-\kappa_b^2\right)\right)}{\left(p^2-\xi_{2\, b} M_b^2\right)}\right)
    \end{pmatrix},
\end{align}\normalsize
where we have defined
\begin{align}\label{eq:def A}
    A(p^2) &=\frac{1}{\left(1-\kappa_b ^2\right)}\frac{p^2-M_b^2}{ p^2-M_b^2/\left(1-\kappa_b ^2\right)}.
\end{align}

For two fields $A_{i}^\mu,\ A_{j}^\nu$, the full two-point function $G_{ij}^{\munu}$ is defined as
\begin{equation}
    \bra{\Omega}T\left\{A_{i}^\mu(x)A_{j}^\nu(y)\right\}\ket{\Omega}=\int\frac{d^4p}{(2\pi)^4}e^{ip(x-y)}iG_{ij}^\munu(p^2),
\end{equation}
and we denote its tree-level contribution by $iP_{ij}^\munu$. Furthermore, let $i\Pi_{ij}^\munu(p^2)$ be the all-order one-particle irreducible (1PI) contribution. Then, suppressing the Lorentz indices $\munu$, the two-point function $G_{ij}$ can be written as
\begin{align}
    iG_{ij}&=\begin{tikzpicture}[baseline=-\the\dimexpr\fontdimen22\textfont2\relax]
    \begin{feynman}
    \vertex (a){\(i\)};
    \vertex [right = of a] (b){\(j\)};
    \diagram* {
	    (a) --[photon] (b),
};
\end{feynman}
\end{tikzpicture}+\begin{tikzpicture}[baseline=-\the\dimexpr\fontdimen22\textfont2\relax]
    \begin{feynman}
    \vertex (a){\(i\)};
    \vertex [right = of a, blob] (b){1PI};
    \vertex [right = of b] (c){\(j\)};
    \diagram* {
	    (a) --[photon] (b) --[photon] (c),
};
\end{feynman}
\end{tikzpicture}+\begin{tikzpicture}[baseline=-\the\dimexpr\fontdimen22\textfont2\relax]
    \begin{feynman}
    \vertex (a){\(i\)};
    \vertex [right = of a, blob] (b){1PI};
    \vertex [right = of b, blob] (c){1PI};
    \vertex [right = of c] (d){\(j\)};
    \diagram* {
	    (a) --[photon] (b) --[photon] (c) --[photon] (d),
};
\end{feynman}
\end{tikzpicture}+\cdots\nonumber\\
&=iP+iP(i\Pi iP)+iP(i\Pi iP)^2+\cdots\nonumber\\
&=iP(1+\Pi P)^{-1}.
\end{align}
Due to gauge invariance, $\Pi_{ij}^{\mu\nu}(p^2)$ is transversal, i.e., $p_\mu\Pi_{ij}^{\mu\nu}(p^2)=~0$. Thus, the Lorentz structure can be factored out, and we may write
\begin{align} \label{eq:def pi mu nu}
    \Pi_{ij}^{\mu\nu}(p^2)=-(p^2g^{\mu\nu}-p^\mu p^\nu)\Pi_{ij}(p^2).
\end{align}
In the next section, we compute the 1-loop UV divergences and renormalize our theory.
\subsection{Renormalization}
To renormalize the theory, we introduce the rescalings
\begin{alignat*}{5}
&\psi_b=\sqrt{Z_2}\psi,\qquad && A_{1\,b}^\mu=\sqrt{Z_3}A_{1}^\mu,\qquad && A_{2\,b}^\mu=\sqrt{Z_4}A_{2}^\mu, \qquad &&m_b=Z_mm,\qquad && e_b=\mu^\epsilon Z_ee,\\
& M_b=\sqrt{Z_M}M,\qquad &&\kappa_b=Z_\kappa\kappa, \qquad && \frac{1}{\xi_{1\,b}}=Z_{\xi_1}\frac{1}{\xi_{1}}, \qquad && \frac{1}{\xi_{2\,b}}=Z_{\xi_2}\frac{1}{\xi_{2}},\qquad &&\phi_b=\sqrt{Z_\phi}\phi.
\end{alignat*}
The renormalization factors are decomposed into counterterms as $Z_i=1+\delta_i$, and we define the two combined renormalization factors
\begin{alignat*}{3}
    Z_1&:=Z_eZ_2\sqrt{Z_3}\qquad &&\Longrightarrow\qquad \delta_1&&:=\delta_e+\delta_2+\frac{1}{2}\delta_3,\\
    Z_\mathrm{mix}&:=Z_\kappa\sqrt{Z_3}\sqrt{Z_2}\qquad&&\Longrightarrow\qquad \delta_\mathrm{mix}&&:=\delta_\kappa+\frac{1}{2}(\delta_3+\delta_4).
\end{alignat*}
The original Lagrangian \eqref{eq:original Lagrangian} can now be written as a "renormalized" Lagrangian plus the counterterm Lagrangian
\begin{align}
\cL_\delta=&-\frac{\delta_3}{4}(F_1^{\mu\nu})^2 +i\delta_2\gamma^\mu\overline{\psi}\partial^\mu\psi-e\delta_1\gamma_\mu\overline{\psi}A_1^\mu\psi-m(\delta_m+\delta_2)\overline{\psi}\psi\nonumber \\
&-\frac{\delta_3+\delta_{\xi_1}}{2\xi_1}(\partial_\mu A_1^\mu)^2
-\kappa\frac{\delta_{\mathrm{mix}}}{2}F_1^{\mu\nu}F_{2\,\mu\nu} 
-\frac{\delta_4}{4}(F_2^{\mu\nu})^2-\frac{\delta_4+\delta_{\xi_2}}{2\xi_2}(\partial_\mu A_2^\mu)^2 \\
&+(\delta_M+\delta_4)\frac{M^2}{2}(A_2^\mu)^2+\frac{\delta_\phi}{2}(\partial_\mu\phi)^2-\xi_2\frac{M^2}{2}(\delta_{\xi_2}+\delta_M+\delta_\phi)\phi^2\nonumber.
\end{align}
The Feynman rules for the counterterm Lagrangian are:
\begin{equation}
\begin{split}
     \begin{tikzpicture}[baseline=-\the\dimexpr\fontdimen22\textfont2\relax,transform shape,scale=0.6]
    \begin{feynman}
    \vertex (a){\(\mu\)};
    \vertex [right = of a, crossed dot] (b){};
    \vertex [right = of b] (c){\(\nu\)};
    \diagram* {
	    (a) --[photon] (b) --[photon] (c),
};
\end{feynman}
\end{tikzpicture} &=-i\delta_3(p^2g^{\mu\nu}-p^\mu p^\nu)-i\frac{\delta_3+\delta_{\xi_1}}{\xi_1}p^\mu p^\nu, \\
    \begin{tikzpicture}[baseline=-\the\dimexpr\fontdimen22\textfont2\relax,transform shape,scale=0.6]
    \begin{feynman}
    \vertex (a){\(\mu\)};
    \vertex [right = of a, crossed dot] (b){};
    \vertex [right = of b] (c){\(\nu\)};
    \diagram* {
	    (a) --[photon,thick] (b) --[photon,thick] (c), (a)--(b)--(c),
};
\end{feynman}
\end{tikzpicture} &=-i\delta_4(p^2g^{\mu\nu}-p^\mu p^\nu)+i(\delta_M+\delta_4)M^2g^{\mu\nu}-i\frac{\delta_4+\delta_{\xi_2}}{\xi_1}p^\mu p^\nu, \\
    \begin{tikzpicture}[baseline=-\the\dimexpr\fontdimen22\textfont2\relax,transform shape,scale=0.6]
    \begin{feynman}
    \vertex (a){\(\mu\)};
    \vertex [right = of a, crossed dot] (b){};
    \vertex [right = of b] (c){\(\nu\)};
    \diagram* {
	    (a) --[photon,thick] (b) --[photon] (c), (a)--(b),
};
\end{feynman}
\end{tikzpicture} &=-i\delta_{\mathrm{mix}}\kappa(p^2g^{\mu\nu}-p^\mu p^\nu), \\
    \begin{tikzpicture}[baseline=-\the\dimexpr\fontdimen22\textfont2\relax,transform shape,scale=0.6]
    \begin{feynman}
    \vertex (a){};
    \vertex [right = of a, crossed dot] (b){};
    \vertex [right = of b] (c){};
    \diagram* {
	    (a) --[fermion] (b) --[fermion] (c),
};
\end{feynman}
\end{tikzpicture} &=i(\slashed{p}\delta_2-m(\delta_m+\delta_2)), \\
    \begin{tikzpicture}[baseline=-\the\dimexpr\fontdimen22\textfont2\relax,transform shape,scale=0.6]
    \begin{feynman}
    \vertex (a){};
    \vertex [right = of a, crossed dot] (b){};
    \vertex [right = of b] (c){};
    \diagram* {
	    (a) --[scalar] (b) --[scalar] (c),
};
\end{feynman}
\end{tikzpicture} &=i(p^2\delta_\phi-\xi_2M^2(\delta_{\xi_2}+\delta_M+\delta_\phi)), \\
    \begin{tikzpicture}[baseline=-\the\dimexpr\fontdimen22\textfont2\relax,transform shape,scale=0.55]
    \begin{feynman}
    \vertex (a){\(\mu\)};
    \vertex [left = of a, crossed dot] (b){};
    \vertex [above left = of b] (c){};
    \vertex [below left = of b] (d){};
    \diagram* {
	    (a) --[photon] (b) --[anti fermion] (c), (b)--[fermion] (d),
};
\end{feynman}
\end{tikzpicture} &=-ie\delta_1\gamma^\mu. \label{eq:counterterms basis1}
\end{split}
\end{equation}
Throughout this paper, for any quantity $f$, we write $f^{[l]}$ for the $l$-loop contribution to $f$. There are no amputated loop-diagrams with external dark photons or Stückelberg fields, so there are only three one-loop diagrams we need to compute; the photon and electron self-energy, and the vertex correction, respectively:
\begin{align*}
    i\Pi_{11}^{[1]\,\mu\nu}(p^2)&= \begin{tikzpicture}[baseline=-\the\dimexpr\fontdimen22\textfont2\relax,transform shape,scale=0.6]
    \begin{feynman}
    \vertex (a);
    \vertex [right = of a] (b);
    \vertex [right = of b] (c);
    \vertex [right = of c] (d);
    \diagram* {
	    (a) --[photon] (b) -- [anti fermion,half left](c) -- [photon](d), (c) -- [anti fermion,half left](b),
};
\end{feynman}
\end{tikzpicture} + 
\begin{tikzpicture}[baseline=-\the\dimexpr\fontdimen22\textfont2\relax,transform shape,scale=0.6]
    \begin{feynman}
    \vertex (a);
    \vertex [right = of a, crossed dot] (b){};
    \vertex [right = of b] (c);
    \diagram* {
	    (a) --[photon] (b) --[photon] (c),
};
\end{feynman}
\end{tikzpicture},\qquad
i\Sigma^{[1]}(\slashed{p}) =
    \begin{tikzpicture}[baseline=-\the\dimexpr\fontdimen22\textfont2\relax,transform shape,scale=0.6]
    \begin{feynman}
    \vertex (a);
    \vertex [right = of a] (b);
    \vertex [right = of b] (c);
    \vertex [right = of c] (d);
    \diagram* {
	    (a) --[fermion] (b) -- [photon,half left](c) -- [fermion](d), (b) -- [fermion](c),
};
\end{feynman}
\end{tikzpicture}+
\begin{tikzpicture}[baseline=-\the\dimexpr\fontdimen22\textfont2\relax,transform shape,scale=0.6]
    \begin{feynman}
    \vertex (a){};
    \vertex [right = of a, crossed dot] (b){};
    \vertex [right = of b] (c){};
    \diagram* {
	    (a) --[fermion] (b) --[fermion] (c),
};
\end{feynman}
\end{tikzpicture}, \\
i\Gamma_1^{[1]\,\mu} (p^2) &=
    \begin{tikzpicture}[baseline=-\the\dimexpr\fontdimen22\textfont2\relax,transform shape,scale=0.55]
    \begin{feynman}
        \vertex at (0, 1) (i1) ;
        \vertex at (0,-1) (i2) ;
        \vertex at (1.5, 1) (a);
        \vertex at (1.5,-1) (b);
        \vertex at (3, 0) (c);
        \vertex at (4.5, 0) (f);
        \diagram*{
            (i1) -- [fermion] (a),
            (i2) -- [anti fermion] (b),
            (a) -- [fermion] (c) -- [fermion] (b),
            (a) -- [photon] (b),
            (c) -- [photon] (f),
        };
    \end{feynman}
\end{tikzpicture}+\begin{tikzpicture}[baseline=-\the\dimexpr\fontdimen22\textfont2\relax,transform shape,scale=0.55]
    \begin{feynman}
    \vertex (a);
    \vertex [left = of a, crossed dot] (b){};
    \vertex [above left = of b] (c){};
    \vertex [below left = of b] (d){};
    \diagram* {
	    (a) --[photon] (b) --[anti fermion] (c), (b)--[fermion] (d),
};
\end{feynman}
\end{tikzpicture}.
\end{align*}
The latter two of these objects differ from the ones in standard QED by the fact that we have to use the modified propagator $iP_{11}$ for the photon, defined through \cref{eq:matrix prop mixed basis}. However, since we only extract the UV poles, the extra factor $A$ simplifies to
\begin{align}
    A(p^2) \xrightarrow[]{p\to \infty}\frac{1}{1-\kappa^2}.
\end{align}
The tensor structures might be simplified, for example, using \texttt{FeynCalc} \cite{Mertig:1990an, Shtabovenko:2023idz}. After a straightforward computation, we find the UV poles
\begin{align}
    i\Pi_{11}^{[1]\,\mu\nu}(p^2)&=\frac{-e^2}{12\pi^2}(p^2g^{\mu\nu}-p^\mu p^\nu)\frac{1}{\epsilon}-i\delta_3(p^2g^{\mu\nu}-p^\mu p^\nu)-i\frac{\delta_3+\delta_{\xi_1}}{\xi_1}p^\mu p^\nu,\nonumber \\
    i\Sigma^{[1]}(\slashed{p}) &= \frac{i e^2 }{16 \pi ^2 \epsilon }\left(\xi _1 (\slashed{p}-m)
    - 3 m\frac{ 1 }{ 1-\kappa ^2 }\right)+i(\slashed{p}\delta_2-m(\delta_m+\delta_2)), \\
    i\Gamma_1^{[1]\,\mu}(p^2) &= -\frac{i e^3 }{16 \pi ^2 \epsilon }\xi _1 \gamma^{\mu }-ie\delta_1\gamma^\mu.\nonumber
\end{align}
From here, we find the minimal subtraction (MS) counterterms at one-loop order,
\begin{equation}
    \begin{aligned}
        \delta^{[1]}_3 &= -\delta^{[1]}_{\xi_1} = -\frac{e^2}{12\pi^2}\frac{1}{\epsilon}, \qquad
    &\delta^{[1]}_2 = -\frac{ e^2 }{16 \pi ^2 \epsilon } \xi_1, \\
   \delta^{[1]}_m &= -\frac{ e^2 }{16 \pi ^2 \epsilon } \frac{ 3 }{ 1-\kappa ^2 }, \qquad 
   &\delta^{[1]}_1 = -\frac{ e^2 }{16 \pi ^2 \epsilon }\xi _1.
    \end{aligned}
\end{equation}
Since there are no one-loop diagrams for the dark-photon self-energy or the mixed photon-dark-photon two-point function, we simply have that the corresponding counterterms in \cref{eq:counterterms basis1} vanish:
\begin{align}
    \delta^{[1]}_4 = \delta^{[1]}_M = \delta^{[1]}_{\xi_2}=0, \qquad
    \delta^{[1]}_{\text{mix}} =0 \quad \implies \quad \delta^{[1]}_\kappa = -\frac{1}{2} \delta^{[1]}_3.
\end{align}
\subsection{Renormalization Group}
Here we derive the RGEs to one-loop order. As usual, we define the beta function for a generic coupling $g$ as $\beta(g):= \mu \dv{\mu} g |_{\epsilon=0}$ and the anomalous dimension of a generic parameter $\Lambda$ as $\gamma_{\Lambda}:= \frac{\mu}{ \Lambda} \dv{\mu} \Lambda|_{\epsilon=0}$. Using that the bare parameters are independent of $\mu$, we get
\begin{align}
    0 = \mu \dv{\mu} e_b =  \mu  \mu^\epsilon e Z_e \left(\epsilon + \frac{\mu}{e}\dv{e}{\mu} + \frac{\mu}{Z_e}\dv{Z_e}{\mu}\right),
\end{align}
and for the $\beta$-function
\begin{align}
    \beta^{[1]}(e) = \frac{e^3}{12\pi^2},
\end{align}
as in standard QED. We also have that
\begin{align}
    0 = \mu \dv{\mu} m_b= Z_m m \left( \frac{\mu}{m}\dv{m}{\mu} + \frac{\mu}{Z_m} \dv{Z_m}{\mu} \right),
\end{align}
so the anomalous dimension for $m$ is
\begin{align}
    \gamma^{[1]}_m =  - \frac{3 e^2 }{8 \pi ^2 } \frac{1}{ 1-\kappa ^2 },
\end{align}
which reproduces the standard QED result upon taking the limit $\kappa\to 0$. Furthermore, we capture the running of $\kappa$ in the quantity
\begin{align}
    \gamma^{[1]}_\kappa =  \frac{ e^2 }{12 \pi^2 }.
\end{align}
Finally, for the gauge fixing parameter, we find
\begin{align}
    \gamma_{\xi_1^{-1}}^{[1]} = \frac{e^2}{6\pi^2}.
\end{align}
The remaining parameters in our Lagrangian do not run at one loop in this basis. The picture changes once we perform a change of basis in field space to remove the mixing term $-\frac{\kappa_b}{2}F_{1\,b}^{\mu\nu}F_{2\,b,\mu\nu}$ in the Lagrangian, which will be the subject of the next section.
\section{Basis Without Kinetic Mixing}\label{sec:basis without kin mix}
We can rotate in field space to trade the mixing term in the Lagrangian \eqref{eq:original Lagrangian} for an explicit dark-photon-fermion coupling and non-diagonal gauge fixing functions. To do so, we start from our original Lagrangian and substitute
\begin{align}
    \begin{pmatrix}
        A_{1\, b}^\mu\\[2pt]
        A_{2\, b}^\mu
    \end{pmatrix}=\begin{pmatrix}
        1&-\frac{\kappa_b}{\sqrt{1-\kappa_b^2}}\\
        0&\frac{1}{\sqrt{1-\kappa_b^2}}
    \end{pmatrix}\cdot\begin{pmatrix}
        \widetilde{A}_{1\, b}^\mu\\[2pt] \widetilde{A}_{2\, b}^\mu
    \end{pmatrix}
\end{align}
to remove the kinetic mixing.
Note that the choice of rotation matrix is not unique. One can always apply an additional orthogonal rotation keeping the kinetic terms canonically normalized, but we do not need this extra degree of freedom here.
Dropping the tilde on the field variables, the original Lagrangian becomes 
\begin{align} \label{eq:Lagrangian basis 2}
    \cL=& -\frac{1}{4}(F_{1\, b}^{\mu\nu})^2
    -\frac{1}{4}( F_{2\, b}^{\mu\nu})^2
    +i\overline{\psi}_b\gamma_\mu\partial^\mu\psi_b -\widetilde m_b\overline{\psi}_b\psi_b\nonumber\\
    &-e_{1\, b} \overline{\psi}_b\gamma_\mu A_{1\, b}^\mu \psi_b 
    - e_{2\, b}\overline{\psi}_b\gamma_\mu A_{2\, b}^\mu \psi_b -\frac{1}{2\widetilde \xi_{1\, b}}\left(\partial_\mu  A_{1\, b}^\mu + \frac{e_{2\, b}}{e_{1\, b}}\partial_\mu  A_{2\, b}^\mu \right)^2 \\
    & -\frac{1}{2\widetilde\xi_{2\, b}}(\partial_\mu \widetilde A_{2\, b}^\mu)^2 +\frac{\widetilde M_b^2}{2}( A_{2\, b}^\mu)^2+\frac{1}{2}(\partial_\mu\phi_b)^2-\widetilde\xi_{2\, b}\frac{\widetilde M_b^2}{2}\phi_b^2,\nonumber
\end{align}
with the new parameters satisfying
\begin{equation}
\begin{aligned}
    \widetilde m_b      &= m_b, 
    &\qquad \widetilde M_b      &= \frac{M_b}{\sqrt{1-\kappa_b^2}}, 
    &\qquad e_{1\, b}   &= e_b, \\[6pt]
    e_{2\, b}           &= -\frac{e_b \kappa_b}{\sqrt{1-\kappa_b^2}},
    &\qquad \widetilde \xi_{1\, b} &= \xi_{1\, b}, 
    &\qquad \widetilde \xi_{2\, b} &= (1-\kappa_b^2)\,\xi_{2\, b}.
\end{aligned}
\label{eq:parameter relations}
\end{equation}
Now we have an explicit electron-dark-photon coupling, but the kinetic mixing is removed. The rotation also produced non-diagonal gauge fixing functions, but we will show explicitly how the off-diagonal contributions cancel when computing any correlation function in the next section.
\subsection{Tree-Level Propagator}\label{sec:basis 2 tree level props}
Due to the non-diagonal gauge fixing function, the matrix-valued gauge boson propagator still has non-zero off-diagonal elements.
The Fourier transformed quadratic terms are \small
\begin{equation}\label{eq: P inv minimal basis}
    P^{-1}:=\begin{pmatrix}
        -p^2\gmunu+\left(1-\frac{1}{\xi_{1\, b}}\right)p_\mu p_\nu & - \frac{e_{2\, b}}{e_{1\,b}}\frac{1}{\xi_{1\, b}}p_\mu p_\nu \\
        - \frac{e_{2\, b}}{e_{1\,b}}\frac{1}{\xi_{1\, b}}p_\mu p_\nu & (-p^2+M_b^2)\gmunu+\left(1+ \frac{e_{2\, b}}{e_{1\,b}} \frac{1}{\xi_{1\, b}}-\frac{1}{\xi_{2\, b}}\right)p_\mu p_\nu
    \end{pmatrix},
\end{equation}\normalsize
and thus we find for the propagator\small
\begin{align}
i P^{\mu \nu}
=\begin{pmatrix}
   -\frac{i}{p^2} \left(g^{\mu \nu} - \left( \frac{ (1-\xi_{1\, b}) }{p^2 } - \frac{e_{2\, b}}{e_{1\, b}} C(p^2)\right) p^{\mu }p^{\nu } \right) & iC(p^2)\frac{ p^{\mu }p^{\nu }}{ p^2} \\
 iC(p^2)\frac{ p^{\mu }p^{\nu }}{ p^2} & -\frac{i}{\left(p^2-M_b^2\right)}\left( g^{\mu \nu} - \frac{(1-\xi_{2\, b}) p^{\mu }p^{\nu }}{ \left(p^2-M_b^2 \xi_{2\, b}\right)}\right)
\end{pmatrix},
\end{align}\normalsize
where 
\begin{align}
    C(p^2)=\xi_{2\, b} \frac{e_{2\, b}}{e_{1\, b}}\frac{1}{ p^2-M_b^2 \xi_{2\, b}}.
\end{align}
We see that, even having removed the $\kappa$-term in the Lagrangian, the off-diagonal elements are still non-vanishing. However, they are unphysical, as indicated by the overall factor of $\xi_{2\, b}$. To see this explicitly in a general gauge, let us define
\begin{align}
    iP^{\mu \nu}_{11} := i\widetilde P^{\mu \nu}_{11} - i\frac{e_{2\, b}}{e_{1\, b}}C(p^2) \frac{p^{\mu }p^{\nu } }{p^2},
\end{align}
i.e., split $P^{\mu \nu}_{11}$ into the usual photon propagator $\widetilde P^{\mu \nu}_{11}$ plus an additional piece.

Now, let us consider an arbitrary Feynman diagram in this theory. For every gauge boson propagator that appears in the diagram, one also has to consider the diagrams obtained by replacing that propagator with either a mixed propagator or the one of the other gauge boson. Summing over these three contributions, taking into account the different couplings of attaching each of these propagators to the rest of the diagram, gives
\begin{align}\label{eq:cancellation of mixed gauge}
    e_{1\, b}^2 P^{\mu \nu}_{11} + e_{1\, b} e_{2\, b} P^{\mu \nu}_{12} + e_{2\, b}^2 P^{\mu \nu}_{22}  = e_{1\, b}^2\widetilde P^{\mu \nu}_{11} + e_{2\, b}^2 P^{\mu \nu}_{22}.
\end{align}
This shows that the off-diagonal part of the gauge-fixing term has no effect and the theory behaves as if we had a massless photon and a massive vector boson without kinetic mixing, i.e., with diagonal tree-level propagators. 
\subsection{UV Poles}
In this basis without the mixed kinetic term, loop corrections still induce a non-vanishing off-diagonal gauge boson two-point function, even though there is no tree-level term in the Lagrangian. Without the parameter $\kappa$, we cannot make the individual off-diagonal two-point functions finite. Instead, as renormalization conditions we impose all observables, i.e. the masses and $S$-matrix elements, to be finite (cf. \cite{Bijnens:2018rqw}). The two-point functions are not physical observables, and thus, they are not required to be finite. Accordingly, we introduce renormalization constants only for the two masses and two couplings of the theory (in the same fashion as in \cite{Bijnens:2018rqw} for the scalar case):
\begin{align}
    \widetilde m_b=Z_{\widetilde m}\widetilde m,\qquad e_{1\, b}= Z_{ e_1} e_1,\qquad \widetilde M_b=\sqrt{ Z_{\widetilde M}}\widetilde M, \qquad e_{2\, b}= Z_{e_2} e_2.
\end{align}
Following \cite{Bijnens:2018rqw}, we do not introduce field rescalings here but, equivalently, we will instead include the residues of the two-point functions explicitly in the LSZ theorem in \cref{sec:lsz}.
As before, we decompose the renormalization factors as $Z_i=1+\delta_i$, and the counterterm Lagrangian then reads
\begin{align}\label{eq: counterterm Lagrangian minimal basis}
\cL_\delta=& 
-e_1\delta_{e_1}\gamma_\mu\overline{\psi} A_1^\mu\psi - e_2 \delta_{e_2}\gamma_\mu\overline{\psi} A_2^\mu\psi
- \widetilde m\delta_{\widetilde m} \overline{\psi}\psi
+ \delta_{\widetilde M} \frac{\widetilde M^2}{2}(A_2^\mu)^2
-\xi_2\frac{\widetilde M^2}{2} \delta_{\widetilde M} \phi^2.
\end{align}
The Feynman rules for the counterterms are
\begin{equation}
\begin{split}
    \begin{tikzpicture}[baseline=-\the\dimexpr\fontdimen22\textfont2\relax,transform shape,scale=0.6]
    \begin{feynman}
    \vertex (a){\(\mu\)};
    \vertex [right = of a, crossed dot] (b){};
    \vertex [right = of b] (c){\(\nu\)};
    \diagram* {
	    (a) --[photon,thick] (b) --[photon,thick] (c), (a)--(b)--(c),
};
\end{feynman}
\end{tikzpicture} &=+i \delta_{\widetilde M} \widetilde M^2 g^{\mu\nu},\quad
    \begin{tikzpicture}[baseline=-\the\dimexpr\fontdimen22\textfont2\relax,transform shape,scale=0.6]
    \begin{feynman}
    \vertex (a){};
    \vertex [right = of a, crossed dot] (b){};
    \vertex [right = of b] (c){};
    \diagram* {
	    (a) --[fermion] (b) --[fermion] (c),
};
\end{feynman}
\end{tikzpicture} =-i \widetilde m\delta_{\widetilde m},\quad
    \begin{tikzpicture}[baseline=-\the\dimexpr\fontdimen22\textfont2\relax,transform shape,scale=0.6]
    \begin{feynman}
    \vertex (a){};
    \vertex [right = of a, crossed dot] (b){};
    \vertex [right = of b] (c){};
    \diagram* {
	    (a) --[scalar] (b) --[scalar] (c),
};
\end{feynman}
\end{tikzpicture} =-i \xi_2 \widetilde M^2 \delta_{\widetilde M}\\
    \begin{tikzpicture}[baseline=-\the\dimexpr\fontdimen22\textfont2\relax,transform shape,scale=0.6]
    \begin{feynman}
    \vertex (a){\(\mu\)};
    \vertex [left = of a, crossed dot] (b){};
    \vertex [above left = of b] (c){};
    \vertex [below left = of b] (d){};
    \diagram* {
	    (a) --[photon] (b) --[anti fermion] (c), (b)--[fermion] (d),
};
\end{feynman}
\end{tikzpicture} &=-ie_1\delta_{e_1}\gamma^\mu,\quad\quad\,\,\,
    \begin{tikzpicture}[baseline=-\the\dimexpr\fontdimen22\textfont2\relax,transform shape,scale=0.6]
    \begin{feynman}
    \vertex (a){\(\mu\)};
    \vertex [left = of a, crossed dot] (b){};
    \vertex [above left = of b] (c){};
    \vertex [below left = of b] (d){};
    \diagram* {
	    (a) --[photon,thick] (b) --[anti fermion] (c), (b)--[fermion] (d), (a) -- (b)
};
\end{feynman}
\end{tikzpicture} = -i e_2\delta_{e_2}\gamma^\mu.
\end{split}
\end{equation}
Next, we will compute the one-loop UV divergences. As we have shown in \cref{sec:basis 2 tree level props}, we may work with the standard diagonal propagators
\begin{equation}
\begin{split}
\begin{tikzpicture}[baseline=-\the\dimexpr\fontdimen22\textfont2\relax]
    \begin{feynman}
    \vertex (a){\(\mu\)};
    \vertex [right = of a] (b){\(\nu\)};
    \diagram* {
	    (a) --[photon] (b),
};
\end{feynman}
\end{tikzpicture} &= \frac{-i}{p^2} \left( \gmunu - (1-\widetilde\xi_1)\frac{p_\mu p_\nu }{p^2} \right),\\
\begin{tikzpicture}[baseline=-\the\dimexpr\fontdimen22\textfont2\relax]
    \begin{feynman}
    \vertex (a){\(\mu\)};
    \vertex [right = of a] (b){\(\nu\)};
    \diagram* {
	    (a) --[photon,thick] (b), 
            (a)--(b),
};
\end{feynman}
\end{tikzpicture} &= \frac{-i}{p^2-\widetilde M^2} \left( \gmunu - (1-\widetilde\xi_2)\frac{p_\mu p_\nu }{p^2-\widetilde \xi_2 \widetilde M^2} \right). \label{eq: standard tree level props}
\end{split}
\end{equation}
The relevant one-loop quantities are four two-point functions, which we denote by
\begin{align*}
    i\Pi_{11}^{[1]\,\mu\nu}&= \begin{tikzpicture}[baseline=-\the\dimexpr\fontdimen22\textfont2\relax,transform shape,scale=0.6]
    \begin{feynman}
    \vertex (a);
    \vertex [right = of a] (b);
    \vertex [right = of b] (c);
    \vertex [right = of c] (d);
    \diagram* {
	    (a) --[photon] (b) -- [anti fermion,half left](c) -- [photon](d), (c) -- [anti fermion,half left](b),
};
\end{feynman}
\end{tikzpicture},\qquad
i\Pi_{12}^{[1]\,\mu\nu}= \begin{tikzpicture}[baseline=-\the\dimexpr\fontdimen22\textfont2\relax,transform shape,scale=0.6]
    \begin{feynman}
    \vertex (a);
    \vertex [right = of a] (b);
    \vertex [right = of b] (c);
    \vertex [right = of c] (d);
    \diagram* {
	    (a) --[photon] (b) -- [anti fermion,half left](c) -- [photon,thick](d), (c) -- [anti fermion,half left](b), (c) -- (d)
};
\end{feynman}
\end{tikzpicture},  \\
i\Pi_{22}^{[1]\,\mu\nu} &= \begin{tikzpicture}[baseline=-\the\dimexpr\fontdimen22\textfont2\relax,transform shape,scale=0.6]
    \begin{feynman}
    \vertex (a);
    \vertex [right = of a] (b);
    \vertex [right = of b] (c);
    \vertex [right = of c] (d);
    \diagram* {
	    (a) --[photon,thick] (b) -- [anti fermion,half left](c) -- [photon,thick](d), (c) -- [anti fermion,half left](b),
        (a) -- (b), (c) -- (d)
};
\end{feynman}
\end{tikzpicture}+\begin{tikzpicture}[baseline=-\the\dimexpr\fontdimen22\textfont2\relax,transform shape,scale=0.6]
    \begin{feynman}
    \vertex (a);
    \vertex [right = of a, crossed dot] (b){};
    \vertex [right = of b] (c);
    \diagram* {
	    (a) --[photon,thick] (b) --[photon,thick] (c), (a)--(b)--(c),
};
\end{feynman}
\end{tikzpicture},\\
i\Sigma^{[1]} &=
    \begin{tikzpicture}[baseline=-\the\dimexpr\fontdimen22\textfont2\relax,transform shape,scale=0.6]
    \begin{feynman}
    \vertex (a);
    \vertex [right = of a] (b);
    \vertex [right = of b] (c);
    \vertex [right = of c] (d);
    \diagram* {
	    (a) --[fermion] (b) -- [photon,half left](c) -- [fermion](d), (b) -- [fermion](c),
};
\end{feynman}
\end{tikzpicture} 
+
    \centering
    \begin{tikzpicture}[baseline=-\the\dimexpr\fontdimen22\textfont2\relax,transform shape,scale=0.6]
    \begin{feynman}
    \vertex (a);
    \vertex [right = of a] (b);
    \vertex [right = of b] (c);
    \vertex [right = of c] (d);
    \diagram* {
	    (a) --[fermion] (b) -- [photon, thick, half left](c) -- [fermion](d), (b) -- [fermion](c), (b) -- [half left](c)
};
\end{feynman}
\end{tikzpicture}+\begin{tikzpicture}[baseline=-\the\dimexpr\fontdimen22\textfont2\relax,transform shape,scale=0.6]
    \begin{feynman}
    \vertex (a){};
    \vertex [right = of a, crossed dot] (b){};
    \vertex [right = of b] (c){};
    \diagram* {
	    (a) --[fermion] (b) --[fermion] (c),
};
\end{feynman}
\end{tikzpicture},
\end{align*}
and the two three-point functions
\begin{align*}
    i\Gamma_1^{[1]\,\mu}&= \begin{tikzpicture}[baseline=-\the\dimexpr\fontdimen22\textfont2\relax,transform shape,scale=0.53]
    \begin{feynman}
        \vertex at (0, 1) (i1) ;
        \vertex at (0,-1) (i2) ;
        \vertex at (1.5, 1) (a);
        \vertex at (1.5,-1) (b);
        \vertex at (3, 0) (c);
        \vertex at (4.5, 0) (f);
        \diagram*{
            (i1) -- [fermion] (a),
            (i2) -- [anti fermion] (b),
            (a) -- [fermion] (c) -- [fermion] (b),
            (a) -- [photon] (b),
            (c) -- [photon] (f),
        };
    \end{feynman}
\end{tikzpicture}\quad+\quad
\begin{tikzpicture}[baseline=-\the\dimexpr\fontdimen22\textfont2\relax,transform shape,scale=0.53]
    \begin{feynman}
        \vertex at (0, 1) (i1) ;
        \vertex at (0,-1) (i2) ;
        \vertex at (1.5, 1) (a);
        \vertex at (1.5,-1) (b);
        \vertex at (3, 0) (c);
        \vertex at (4.5, 0) (f);
        \diagram*{
            (i1) -- [fermion] (a),
            (i2) -- [anti fermion] (b),
            (a) -- [fermion] (c) -- [fermion] (b),
            (a) -- [photon,thick] (b),
            (a) -- (b),
            (c) -- [photon] (f),
        };
    \end{feynman}
\end{tikzpicture} \quad+\quad 
\begin{tikzpicture}[baseline=-\the\dimexpr\fontdimen22\textfont2\relax,transform shape,scale=0.6]
    \begin{feynman}
    \vertex (a);
    \vertex [left = of a, crossed dot] (b){};
    \vertex [above left = of b] (c){};
    \vertex [below left = of b] (d){};
    \diagram* {
	    (a) --[photon] (b) --[anti fermion] (c), (b)--[fermion] (d),
};
\end{feynman}
\end{tikzpicture}, \\
i\Gamma_{2}^{[1]\,\mu}&= \begin{tikzpicture}[baseline=-\the\dimexpr\fontdimen22\textfont2\relax,transform shape,scale=0.53]
    \begin{feynman}
        \vertex at (0, 1) (i1) ;
        \vertex at (0,-1) (i2) ;
        \vertex at (1.5, 1) (a);
        \vertex at (1.5,-1) (b);
        \vertex at (3, 0) (c);
        \vertex at (4.5, 0) (f);
        \diagram*{
            (i1) -- [fermion] (a),
            (i2) -- [anti fermion] (b),
            (a) -- [fermion] (c) -- [fermion] (b),
            (a) -- [photon] (b),
            (c) -- [photon, thick] (f),
            (c) -- (f),
        };
    \end{feynman}
\end{tikzpicture}\quad+\quad
\begin{tikzpicture}[baseline=-\the\dimexpr\fontdimen22\textfont2\relax,transform shape,scale=0.53]
    \begin{feynman}
        \vertex at (0, 1) (i1) ;
        \vertex at (0,-1) (i2) ;
        \vertex at (1.5, 1) (a);
        \vertex at (1.5,-1) (b);
        \vertex at (3, 0) (c);
        \vertex at (4.5, 0) (f);
        \diagram*{
            (i1) -- [fermion] (a),
            (i2) -- [anti fermion] (b),
            (a) -- [fermion] (c) -- [fermion] (b),
            (a) -- [photon,thick] (b),
            (a) -- (b),
            (c) -- [photon, thick] (f),
            (c) -- (f),
        };
    \end{feynman}
\end{tikzpicture} \quad+\quad 
\begin{tikzpicture}[baseline=-\the\dimexpr\fontdimen22\textfont2\relax,transform shape,scale=0.6]
    \begin{feynman}
    \vertex (a);
    \vertex [left = of a, crossed dot] (b){};
    \vertex [above left = of b] (c){};
    \vertex [below left = of b] (d){};
    \diagram* {
	    (a) --[photon,thick] (b) --[anti fermion] (c), (b)--[fermion] (d), (a) -- (b)
};
\end{feynman}
\end{tikzpicture}.
\end{align*}
Many of these objects only differ by constants originating from the vertex factors. We find the UV poles
\begin{equation}
\begin{split}
    i\Pi_{11}^{[1]\,\mu\nu}&=\frac{-i e_1^2}{12\pi^2}(p^2g^{\mu\nu}-p^\mu p^\nu)\frac{1}{\epsilon}, \qquad 
    i\Pi_{12}^{[1]\,\mu\nu}=-\frac{i e_1 e_2}{12\pi^2}(p^2g^{\mu\nu}-p^\mu p^\nu)\frac{1}{\epsilon},\\
    i\Pi_{22}^{[1]\,\mu\nu}&=\frac{-i  e_2^2}{12\pi^2}(p^2g^{\mu\nu}-p^\mu p^\nu)\frac{1}{\epsilon}+i \delta_{\widetilde M} \widetilde M^2 g^{\mu\nu},  \\
    i\Sigma^{[1]} &= \frac{i }{16 \pi^2  \epsilon}
    \left( \left(e_1^2\widetilde \xi _1 +e_2^2 \widetilde \xi_2\right)\slashed{p} - \left( e_1^2(\widetilde \xi_1+3)+e_2^2(\widetilde \xi_2+3) \right) \widetilde m\right) -i \widetilde m\delta_{\widetilde m}, \\
    i\Gamma_1^{[1]\,\mu} &=-\frac{i e_1}{16 \pi ^2 \epsilon } \gamma^{\mu } (\widetilde \xi_1 e_1^2+\widetilde \xi_2 e_2^2)-ie_1\delta_{e_1}\gamma^\mu, \\
    i\Gamma_2^{[1]\,\mu}&=-\frac{i e_2 }{16 \pi ^2 \epsilon } \gamma^{\mu } (\widetilde \xi_1 e_1^2+\widetilde \xi_2 e_2^2)-ie_2\delta_{e_2}\gamma^\mu.
\end{split}
\end{equation}
\subsection{Resummed Green's Functions, LSZ Theorem and Renormalization}\label{sec:lsz}
The crucial observation is that in this basis, even if we were to renormalize all parameters, there is no counterterm able to absorb the divergence of $\Pi_{12}^{[1]\,\mu\nu}$, so the off-diagonal element of the full Green's function, $G_{12}$, cannot be made finite. This is not a problem as it is only the poles of $G_{ij}$ (i.e. the masses of the particles) that are observables and in need of renormalization. The other observables that have to be finite are the $S$-matrix elements calculated from the LSZ theorem, which differ from amputated, renormalized Green's functions due to the mixing on the external legs. After accounting for this \cite{Amoros:2001cp}, the $S$-matrix elements may receive a contribution from $G_{12}$. However, this divergence can be absorbed by the coupling counterterms.

The physical masses of the photon and dark photon are the poles of the matrix-valued two-point function $G^{\mu\nu}$. They may be found as the zeros of the determinant of the inverse of $G^{\mu\nu}$ \cite{Amoros:2001cp}:
\begin{equation}\label{eq: det Ginv}
    \det\left(G^{-1}\right)=\det\left((1+\Pi P)P^{-1}\right)=\det\left(P^{-1}+\Pi\right),
\end{equation}
where we again suppressed the Lorentz indices. This allows us to determine the poles without explicitly resumming all 1PI contributions. The determinant evaluates to\small
\begin{equation}
\begin{split}
    \det(G^{-1})=\frac{(p^2)^4}{\widetilde \xi_{1}\widetilde \xi_{2}}\left(p^2\Delta-\widetilde M^2(1+\Pi_{11})\right)^3\left(p^2-\widetilde{\xi}_{2}\widetilde{M}^2\right),\quad
     \Delta =\det\begin{pmatrix}
        \Pi_{11}+1 & \Pi_{12}\\
        \Pi_{12} & \Pi_{22}+1
    \end{pmatrix}.
    \end{split}
\end{equation}\normalsize
The two gauge-independent pole masses $\widetilde m_{\gamma,P}^2$ and $\widetilde M_P^2$ are, respectively, the solutions of $p^2=0$ (i.e. the photon mass) and $\widetilde M^2(1+\Pi_{11})-p^2\Delta=0$, which determines the mass of the dark photon.

As our propagators have a non-standard structure, we also resum the 1PI diagrams to provide the explicit results for the resummed Green's functions in general $R_\xi$-gauge. Note that also in this basis, $\Pi_{ij}^{\mu\nu}$ is transversal due to gauge invariance. In standard QED, the resummed photon two-point function is
\begin{align}
    iG^{\mu\nu}&=\frac{-i}{p^2(1+\Pi)}\left(g^{\mu\nu}-\frac{(1-(1+\Pi)\xi)p^\mu p^\nu}{p^2}\right).
\end{align}
Once we add the dark photon, the expressions become slightly more involved and matrix-valued. In this basis, the components of the resummed Green's function are
\begin{align}
    G_{11}^{\mu\nu}&=\frac{-1}{p^2}\left(\frac{p^2(1+\Pi_{22})-\widetilde M^2}{\Delta p^2-\widetilde M^2(1+\Pi_{11})}\left(g^{\munu}-\frac{p^\mu p^\nu}{p^2}\right) +\left(\tilde\xi_1+\frac{e_{2}^2}{e_{1}^2}\frac{p^2\widetilde{\xi}_2}{p^2-\widetilde{\xi}_2\widetilde{M}^2}\right)\frac{p^\mu p^\nu}{p^2}\right), \nonumber\\
    G_{12}^{\mu\nu}&=\frac{\Pi_{12}}{\Delta p^2-\widetilde M^2(1+\Pi_{11})}\left(g^{\mu\nu}-\frac{p^\mu p^\nu}{p^2}\right)+\frac{e_{2}}{e_{1}}\frac{\widetilde \xi_2}{p^2-\widetilde \xi_2 \widetilde M^2}\frac{p^\mu p^\nu}{p^2},\\
    G_{22}^{\mu\nu}&=\frac{-1}{\Delta p^2-\widetilde M^2(1+\Pi_{11})}
    \left((1+\Pi_{11})\left(g^{\mu\nu}-\frac{p^\mu p^\nu}{p^2-\widetilde \xi_2 \widetilde M^2}\right)+ \widetilde \xi_2\Delta \frac{p^\mu p^\nu }{p^2-\widetilde \xi_2 \widetilde M^2}\right).\nonumber
\end{align}
Note that the poles coincide with the zeros of $\det(G^{-1})=0$. Furthermore, note that, even though the kinetic mixing is removed in this basis, these results are modified due to the presence of the off-diagonal gauge fixing terms in the tree-level propagator. However, just as at tree-level, there is a cancellation of the same type as in \cref{eq:cancellation of mixed gauge} also here, on the level of the resummed Green's function.

For the pole mass $\widetilde M_P^2$ up to one-loop order, we find
\begin{align}
    (\widetilde M_{P}^2)^{[0]}+(\widetilde M_{P}^2)^{[1]} = \widetilde M^2 \left(1-\frac{ e_2^2}{12\pi^2}\frac{1}{\epsilon}+\delta^{[1]}_{\widetilde M}\right),
\end{align}
and thus,
\begin{align}
    \delta^{[1]}_{\widetilde M} = \frac{ e_2^2}{12\pi^2}\frac{1}{\epsilon}.
\end{align}
For the electron, the resummed two-point function is
\begin{align}
    G_e(\slashed{p}) = \frac{1}{\slashed{p}-\widetilde m+\Sigma(\slashed{p})} = \frac{1}{\slashed{p}(1+\Sigma_V)-\widetilde m(1-\Sigma_S)}, \qquad \Sigma(\slashed{p}) = \slashed{p} \Sigma_V + \widetilde m \Sigma_S.
\end{align}
We require the physical pole mass to one-loop order,
\begin{align}
    \widetilde m^{[0]}_P+\widetilde m^{[1]}_P = \widetilde m (1+\delta^{[1]}_{\widetilde m} -\Sigma^{[1]}_S-\Sigma^{[1]}_V),
\end{align}
to be finite, which fixes the counterterm
\begin{align}
    \delta^{[1]}_{\widetilde m} = \Sigma^{[1]}_V+\Sigma^{[1]}_S = - \frac{3}{16\pi^2 \epsilon}(e_1^2+e_2^2).
\end{align}
At this pole we define the residue
\begin{align}
    Z_e = \left( \lim\limits_{\slashed{p}\to \widetilde m_P} (\slashed{p}-\widetilde m_P) G_e(\slashed{p})\right)=\left.\frac{1}{1+ \Sigma_V}\right|_{\,\slashed{p}=\widetilde m_P}.
\end{align}
The K\"all\'en-Lehmann representation for vector bosons in covariant gauge may be written as \cite{Itzykson:1980rh}\small
\begin{equation}
-G_{ij}(p^2)= Z_{ij}^{(k)}\left(\frac{g^\munu-p^\mu p^\nu/M_k^2}{p^2-M_k^2+i\varepsilon} +z_{ij}^{(k)}\frac{p^\mu p^\nu/M_k^2}{p^2-\xi M_k^2+i\varepsilon}\right)+\int_{M_0^2}^\infty\rho_{ij}(s)\frac{g^\munu-p^\mu p^\nu/s}{p^2-s+i\varepsilon}ds,
\end{equation}\normalsize
where $Z_{ij}^{(k)}$ and $z_{ij}^{(k)}$ are scalar numbers and $\rho_{ij}(s)$ a positive measure. The factor $Z_{ij}^{(k)}$ is simply the residue at the pole of the $ij$-two-point function at the state with mass $M_k^2$. For practical purposes it is easiest to find the residue by fixing the gauge to Feynman gauge, $\widetilde{\xi}_1\to0$, $\widetilde{\xi}_2 \to 0$. In particular, we find
\begin{equation}
    \begin{aligned}
            Z_{11}^{(1)}&=\lim\limits_{p^2\to 0}p^2 \left( \frac{1}{p^2}\frac{(p^2(1+\Pi_{22})-\widetilde{M}^2)}{\Delta p^2-\widetilde{M}^2(1+\Pi_{11})} \right)=\left. \frac{1}{1+\Pi_{11}}\right|_{p^2=0}, \\
    Z_{22}^{(2)}&= \lim\limits_{p^2\to \widetilde{M}_P^2}(p^2-\widetilde{M}_P^2)\frac{(1+\Pi_{11})}{\Delta p^2-\widetilde{M}^2(1+\Pi_{11})}=\left.\frac{1+\Pi_{11}}{\frac{\mathrm{d}}{\mathrm{d}p^2}(\Delta p^2-\widetilde{M}^2(1+\Pi_{11}))}\right|_{p^2=\widetilde{M}_P^2},\\
    Z_{12}^{(1)}&=\lim_{p^2\to 0}p^2\left(\frac{-\Pi_{12}}{\Delta p^2-\widetilde{M}^2(1+\Pi_{11})}\right)=0,\\
    Z_{12}^{(2)}&=\lim_{p^2\to \widetilde{M}_P^2}(p^2-\widetilde{M}_P^2)\left(\frac{-\Pi_{12}}{\Delta p^2-\widetilde{M}^2(1+\Pi_{11})}\right)=\left.\frac{-\Pi_{12}}{\frac{\mathrm{d}}{\mathrm{d}p^2}(\Delta p^2-\widetilde{M}^2(1+\Pi_{11}))}\right|_{p^2=\widetilde{M}_P^2}.
    \end{aligned}
\end{equation}
We denote the two physical three-point amplitudes for $e\bar e \gamma$ and $e\bar e \gamma_{\text{dark}}$ by $\mathcal{A}_1$ and $\mathcal{A}_2$, respectively. They may be computed from the LSZ theorem after accounting for mixing on the external legs, as was demonstrated for a scalar theory in \cite{Amoros:2001cp}. In our case, the $S$-matrix elements are then given by
\begin{align}
    \mathcal{A}_i^\mu &= \left( \lim\limits_{\slashed{p}\to \widetilde m_P} (\slashed{p}-\widetilde m_P) G_e(\slashed{p})\right)^2
    \lim\limits_{p^2\to \widetilde m_i^2} (p^2-\widetilde m_i^2) G_{i j }^{\mu \nu}(p^2) \frac{1}{\sqrt{Z_e^2 Z_{ii}^{(i)}}} \Gamma_{j, \nu}\nonumber
    \\ 
    &= \left( \lim\limits_{\slashed{p}\to \widetilde m_P} (\slashed{p}-\widetilde m_P) G_e(\slashed{p})\right)\lim\limits_{p^2\to \widetilde m_i^2} (p^2-\widetilde m_i^2) G_{i i }^{\mu \nu}(p^2) \frac{1}{\sqrt{Z_{ii}^{(i)}}} \Gamma_{i, \nu} \\
    &+
    \left( \lim\limits_{\slashed{p}\to \widetilde m_P} (\slashed{p}-\widetilde m_P) G_e(\slashed{p})\right)\left(\lim\limits_{p^2\to \widetilde m_i^2} (p^2-\widetilde m_i^2) G_{i j }^{\mu \nu}(p^2) \frac{1}{\sqrt{Z_{ii}^{(i)}}} \Gamma_{j, \nu}\right)\Bigg|_{i \neq j},\nonumber
\end{align}
with an implicit sum over the repeated index $j$. Here, we separated the contribution from attaching diagonal or off-diagonal gauge boson legs. 
The divergent piece of the amplitude with an external photon at one loop is given by
\begin{equation}
    \cA_1^{[1]\,\mu}\sim -e_1\gamma^\mu\left(\delta_{e_1}^{[1]}-\frac{e_1^2}{24\pi^2}\frac{1}{\epsilon}\right)\implies \delta^{[1]}_{e_1} = \frac{ e_{1}^2}{24\pi^2}\frac{1}{\epsilon}.
\end{equation}
Similarly, for the dark photon amplitude, we find
\begin{equation}
    \cA_2^{[1]\,\mu}\sim -e_2\gamma^\mu\left(\delta_{e_2}^{[1]}-\frac{e_2^2}{24\pi^2}\frac{1}{\epsilon}-\frac{e_1^2}{12\pi^2}\frac{1}{\epsilon}\right)\implies  \delta^{[1]}_{e_2} = \frac{ e_{1}^2}{12\pi^2 \epsilon} + \frac{  e_{2}^2}{24\pi^2\epsilon}.
\end{equation}
Note that the extra contribution in $\delta^{[1]}_{e_2}$ proportional to $e_1^2$ comes from the fact that we also need to absorb the divergence originating from $G_{12}$.
Next, we study the RGEs for all physical parameters in this basis and then compare them to the results in the basis with kinetic mixing.
\subsection{Renormalization Group}
Again, we give the RGEs here to one-loop order. For $e_1$, we find the same as in standard QED,
\begin{align}
    \beta^{[1]}(e_1)  = \frac{e_1^3}{12\pi^2}.
\end{align}
However, for $e_2$ we have
\begin{align}
    \beta^{[1]}(e_2)
    = - e_2 \left(\epsilon + \frac{\mu}{Z_{e_2}} \left(\dv{Z_{e_2}}{e_1} \dv{e_1}{\mu}+\dv{Z_{e_2}}{e_2} \dv{e_2}{\mu}\right)\right) = \frac{e_1^2 e_2}{6\pi^2}+ \frac{e_2^3}{12\pi^2} .
\end{align}
The anomalous dimension for $\widetilde m$ is
\begin{align}
    \gamma_{\widetilde m}^{[1]} = - \frac{\mu}{Z_{\widetilde m}} \dv{Z_{\widetilde m}}{\mu} = - \frac{\mu}{Z_{\widetilde m}} \left( \dv{Z_{\widetilde m}}{e_1}  \dv{e_1}{\mu} +\dv{Z_{\widetilde m}}{e_2}  \dv{e_2}{\mu}\right) 
    =-\frac{ 3 }{8 \pi ^2} (e_1^2+e_2^2).
\end{align}
In this basis, we also have an anomalous dimension for the dark photon mass $\widetilde M$,
\begin{align}
    \gamma_{\widetilde M}^{[1]} = - \frac{\mu}{\sqrt{Z_{\widetilde M}}} \dv{\sqrt{Z_{\widetilde M}}}{\mu} = - \frac{1}{2} \dv{\delta_{\widetilde M}}{e_2} \mu \dv{e_2}{\mu} = \frac{e_2^2}{12\pi^2} .
\end{align}
\section{Comparison of the Two Bases}\label{sec: comparison}
For the convenience of the reader, we give again the full set of all non-vanishing RGEs in the two bases. In the basis with kinetic mixing, we found
\begin{align}
    \beta^{[1]}(e) =  \frac{e^3}{12\pi^2}, \qquad
    \gamma_m^{[1]}  = - \frac{3 e^2 }{8 \pi ^2 } \frac{1}{ 1-\kappa ^2 } ,\qquad \gamma_\kappa^{[1]} =  \frac{ e^2 }{12 \pi^2 }, \qquad
    \gamma_{\xi_1^{-1}}^{[1]} = \frac{e^2}{6\pi^2}.
\end{align}
In the basis without kinetic mixing, we have
\begin{equation}
    \beta^{[1]}(e_1) = \frac{e_1^3}{12\pi^2}, \quad
    \beta^{[1]}(e_2) =  \frac{e_1^2 e_2}{6\pi^2}+ \frac{e_2^3}{12\pi^2}, \quad
    \gamma_{\widetilde m}^{[1]} = -\frac{ 3 }{8 \pi ^2} (e_1^2+e_2^2), \quad
     \gamma_{\widetilde M}^{[1]} = \frac{e_2^2}{12\pi^2}.
\end{equation} 
The relations between the parameters in the two bases have the same form for the renormalized parameters as for the bare ones. Thus, according to \cref{eq:parameter relations}, we have
\begin{equation}
\begin{split}
    \widetilde m &= m, \quad \widetilde M = \frac{M}{\sqrt{1-\kappa^2}}, \quad e_{1}=e, \quad e_{2} = - \frac{e \kappa}{\sqrt{1-\kappa^2}}.
    \end{split}
\end{equation}
Here we give only the relations for the physical parameters, since we will be interested in matching their RGEs only.

The matching of $e$ with $e_1$ and of $m$ with $\widetilde m$ is trivial: Expressing $e_1$ and $e_2$ on the right-hand side of the RGEs in terms of $e$ and $\kappa$, we immediately find that
\begin{align}
\beta^{[1]}(e) = \beta^{[1]}(e_1), \qquad \gamma_m^{[1]} = \gamma_{\widetilde m}^{[1]}.
\end{align}
To match the remaining physical parameters, we first differentiate the relations of the parameters in the two bases,
\begin{align}\label{eq:for gamma tilde M}
    \gamma_{\widetilde M}^{[1]} = \frac{\mu}{\widetilde M} \dv{\mu} \widetilde M = \sqrt{1-\kappa^2}\frac{\mu}{ M}\dv{\mu} \left(\frac{M}{\sqrt{1-\kappa^2}}\right) = \gamma_M^{[1]} + \sqrt{1-\kappa^2} \mu \dv{\mu}\frac{1}{\sqrt{1-\kappa^2}},
\end{align}
and
\begin{align}\label{eq:for beta e2}
    \beta^{[1]}(e_2) = \mu \dv{\mu} e_2 = \mu \dv{\mu} \left( - \frac{e \kappa}{\sqrt{1-\kappa^2}}\right) = - \frac{ \kappa}{\sqrt{1-\kappa^2}} \beta^{[1]}(e) - e \mu \dv{\mu}\frac{ \kappa}{\sqrt{1-\kappa^2}}.
\end{align}
From \cref{eq:for gamma tilde M} we have
\begin{align}
    \sqrt{1-\kappa^2} \mu \dv{\mu}\frac{1}{\sqrt{1-\kappa^2}} = \gamma_{\widetilde M}^{[1]}-\gamma_{ M}^{[1]},
\end{align}
and with the chain rule, we find
\begin{align}
    \frac{\kappa}{1-\kappa^2} \mu \dv{\kappa}{\mu} = \gamma_{\widetilde M}^{[1]}-\gamma_{ M}^{[1]} = \frac{e_2^2}{12\pi^2} \implies \gamma_\kappa=\frac{e^2}{12\pi^2}.
\end{align}
In the same fashion, from \cref{eq:for beta e2} we get
\begin{align}
    \mu \dv{\mu}\frac{ \kappa}{\sqrt{1-\kappa^2}} = - \frac{1}{e}\left( \frac{ \kappa}{\sqrt{1-\kappa^2}} \beta(e) + \beta(e_2)\right),
\end{align}
and thus
\begin{align}
    \frac{ 1}{(1-\kappa^2)^{3/2}}\mu \dv{\kappa}{\mu} = - \frac{1}{e}\left( \frac{ \kappa}{\sqrt{1-\kappa^2}} \frac{e^3}{12\pi^2} + \frac{e_1^2 e_2}{6\pi^2}+ \frac{e_2^3}{12\pi^2} \right)\implies\gamma_\kappa=  \frac{e^2}{12\pi^2}. 
\end{align}
This shows that the RGEs of all physical parameters in the two bases can be matched consistently with a rotation matrix that runs itself.
\section{Conclusions}\label{sec: conclusions}
We have shown at one-loop order that in a theory with two $U(1)$ gauge bosons, one can choose a renormalization scheme without kinetic mixing. This means that after the kinetic mixing has been rotated away, it is not necessary to reintroduce this parameter to absorb divergences of off-diagonal two-point functions. We showed this by explicitly matching two renormalization schemes for two different bases in field space: one where an additional $U(1)$ gauge boson (dark photon) was added to QED with kinetic mixing and one where this mixing was rotated away. The calculations are done in general covariant $R_\xi$-gauge, and naively, the gauge fixing function obstructs diagonalizability of the tree-level propagator. However, this obstruction is explicitly gauge-dependent, and we have shown that it drops out at each loop order once all contributing diagrams are summed. 

Importantly, in the second scheme, we only renormalize the physical observables, i.e. the masses and scattering amplitudes, and we show that the two schemes can be matched using scale-dependent field redefinitions.

For future work, we plan to extend this analysis to two loops and beyond. At this level, the physical principles described here imply non-trivial cancellations between sub-diagrams to ensure renormalizability. Elucidating this cancellation and proving it to all loop orders would be very illuminating.
\section*{Acknowledgments}
The authors would like to thank Jonas Frerick, Erik Panzer, and Lorenzo Tancredi for insightful discussions during the preparation of this paper.

FT is funded by Royal Society grant number URF\textbackslash R1\textbackslash 201473 and gratefully acknowledges this support. The research of FT is also funded by the European Union ERC Synergy Grant MaScAmp 101167287. Views and opinions expressed are however those of the author(s) only and do not necessarily reflect those of the European Union or the European Research Council. Neither
the European Union nor the granting authority can be held responsible for them.

FF is supported by the research unit FOR 5582 “Modern Foundations of Scattering Amplitudes” funded by the Deutsche Forschungsgemeinschaft
(DFG, German Research Foundation) — Projektnummer 508889767.
\bibliographystyle{JHEP}
\bibliography{refs.bib}
\end{document}